\documentclass[twocolumn,showpacs,preprintnumbers,amsmath,amssymb]{revtex4} 
 
\usepackage{graphicx} 
\usepackage{dcolumn} 
\usepackage{bm} 
 
\begin{document}  
 
\preprint{} 
\input{epsf.tex} 
 
 
\title{Bright and dark excitons in an atom--pair filled optical lattice within a cavity} 
\author{Hashem Zoubi, and Helmut Ritsch} 
 
\affiliation{Institut fur Theoretische Physik, Universitat Innsbruck, Technikerstrasse 25, A-6020 Innsbruck, Austria}  
  
\date{20 June, 2007} 
 
\begin{abstract} 
We study electronic excitations of a degenerate gas of atoms trapped in pairs in an optical lattice. Local dipole-dipole interactions produce a long lived antisymmetric and a short lived symmetric superposition of individual atomic excitations as the lowest internal on-site excitations. Due to the much larger dipole moment the symmetric states couple efficiently to neighbouring lattice sites and can be well represented by Frenkel excitons, while the antisymmetric dark states stay localized. Within a cavity only symmetric states couple to cavity photons inducing long range interactions to form polaritons. We calculate their dispersion curves as well as cavity transmission and reflection spectra to observe them. For a lattice with aspherical sites bright and dark states get mixed and their relative excitation energies depend on photon polarizations. The system should allow to study new types of solid state phenomena in atom filled optical lattices. 
\end{abstract} 
 
 
\maketitle 
 
A dilute gas of bosonic atoms near $T=0$ in an optical-lattice has proved an ideal test-system to study important quantum phenomena of solid state physics with well controllable parameters \cite{Jaksch,Zoller}. A striking example is a reversible quantum phase transition from the superfluid into the Mott-insulator phase \cite{Bloch} simply by changing the optical potential depth. In the Mott insulator case each optical-lattice site is filled with a fixed number of atoms down to one or two atoms per site. For a deep enough lattice the atoms cannot move and form an artificial crystal. Naturally it is now interesting to study further complex solid state phenomena in this system, e.g. by exploiting the internal atomic level structure, which bears a strong analogy to excitonic dynamics of molecular crystals (Frenkel excitons) as predicted in Ref. \cite{Hashem}. By the help of an optical cavity these excitons get strongly coupled over large distances via photons and form polaritons. In the present paper we investigate a special interesting case of such excitons \cite{Davydov} and cavity polaritons \cite{Kavokin} which can appear only for lattice filled with two atoms per site, which is a straight forward to prepare in optical lattices by the help of a Mott insulator state with filling factor 2.    
 
Let us start from a degenerate gas of effective two-level atoms trapped in a $2D$ optical lattice, located within a cavity with a single cavity mode close to resonance with an internal atomic transition. The lattice laser is tuned far off resonance to the atomic excitations and results in light shifts of the ground and excited states with periodicity of half the laser wave length. Here we assume the two optical-lattices for ground and excited states located at the same positions, which can be realized for Alkali or Alkaline atoms. At certain {\it magic} laser frequencies the excited state even experiences an equal shift as the ground state \cite{Zoller,Katori}. We believe that more general lattice configurations might imply new physics, but this goes beyond our aim here. At temperatures close to $T=0$, the atomic center of mass motion is confined to the lowest Bloch band and the ground and excited state atoms can be mathematically treated as two kinds of bosons leading to a Bose-Hubbard model with two bosons. It possesses a rich phase diagram containing a Mott-insulator and a superfluid \cite{Chen}. 

\ 
 
For a deep lattice atomic tunneling is suppressed so that each atom retains his identity with negligible overlap of the electronic wave functions of atoms at different sites. As the ground state wave-function (Wannier function) at each site is still much larger than the size of an atom, we can for the moment also neglect the short range part of the molecular potentials and treat their interaction by a pseudo-potential involving only the scattering length. This makes the artificial atomic lattices similar to Nobel atom or molecular crystals. While the electronic excitations can transfer between atoms due to dipole-dipole interactions, there is no direct electron exchange. A simplified model atomic Hamiltonian then reads \cite{Zoubi}: 
\begin{equation} 
H=\sum_{i,\alpha}\hbar\omega_A\ B_i^{\alpha\dagger}B_i^{\alpha}+\sum_{ij,\alpha,\beta}\hbar J_{ij}^{\alpha\beta}\ B_i^{\alpha\dagger}B_j^{\beta}. 
\end{equation} 
where $B_i^{\alpha\dagger}$ and $B_i^{\alpha}$ are the creation and annihilation operators of an excitation at atom $(i,\alpha)$, respectively. The summation $i$ runs over the lattice sites, while $\alpha$ labels the two atoms at one site. The first part represents local excitation with a transition frequency $\omega_A$, while the second part generates the energy transfer between atoms $(i,\alpha)$ and $(j,\beta)$ with coupling amplitude $J_{ij}^{\alpha\beta}$. In principle $B_i^{\alpha}$ are two-level transition operators that forbid two excitations of the same atom. Nevertheless at low excitation density we neglect the possibility of two excitations on the same atom and we can assume the excitations to behave as Bosons \cite{ZoubiA}. 

As simple example to discuss the basic physics we consider a cubic optical-lattice with energy transfer only between atoms on the same site with amplitude $J^{\alpha\beta}(R\ll a)=-J_0$, and nearest neighbor sites with strength $J^{\alpha\beta}(R=a)=-J_1^{\alpha\beta}$. The on-site part of the Hamiltonian in principle can be diagonalized changing to symmetric and antisymmetric entangled excitations $B^1_i=\frac{B_i^s+B_i^a}{\sqrt{2}}$, and $B^2_i=\frac{B_i^s-B_i^a}{\sqrt{2}}$, to get 
\begin{eqnarray} 
H&=&-\sum_{\langle i\neq j\rangle}\hbar\left\{ (J_1+J'_1)\ B_i^{s\dagger}B_j^{s}+(J_1-J'_1)\ B_i^{a\dagger}B_j^{a}\right\} \nonumber \\ 
&+&\sum_i\hbar\left(\omega_a\ B_i^{a\dagger}B_i^{a}+\omega_s\ B_i^{s\dagger}B_i^{s}\right), 
\end{eqnarray}  
where $\omega_{s}=\omega_A-J_0$ and $\omega_{a}=\omega_A+J_0$.  Note that evaluating the local coupling requires integration of dipole-dipole coupling over the local wave-functions but will eventually just give a fixed parameters. For these energy transfer parameters we assumed $J_1^{11}=J_1^{22}=J_1$ and $J_1^{12}=J_1^{21}=J'_1$. As $J_1\approx J'_1$, the nonlocal energy transfer is large only for the symmetric excitations which decay with line-width of $\hbar\Gamma_s\approx 2\hbar\Gamma$, which is two times the atomic line-width \cite{Ficek}. In contrast the antisymmetric states are shown to be metastable with a long radiative life time and negligible next neighbor coupling. 
 
Using the lattice symmetry a delocalized excitation can be represented in quasi-momentum space by a propagating wave with wave vector ${\bf k}$. Such a quasi-particle is called Frenkel-exciton in molecular crystals \cite{Davydov,Zoubi} and our Hamiltonian can be diagonalized by help of these quasi-particles $B_i^{\nu}=\frac{1}{\sqrt{N}}\sum_{\bf k}B^{\nu}_{\bf k}\ e^{i{\bf k}\cdot{\bf r}_i}$. Here $(\nu=s,a)$, $N$ is the number of lattice sites, and ${\bf r}_i$ is the position of site $i$.  In $2D$ optical lattice ${\bf k}$ takes the values ${\bf k}=(k_x,k_y)=\frac{2\pi}{Ma}\left(n_x,\ n_y\right)$, and $n_{x,y}=0,\pm 1,\cdots,\pm \frac{M}{2}$, with $N=M\times M$, and $M$ is an even number. The Hamiltonian casts into $H=\sum_{{\bf k}\nu}\hbar\omega_{\nu}({\bf k})\ B_{\bf k}^{\nu\dagger}B_{\bf k}^{\nu}$, with the energy dispersion for the symmetric and antisymmetric branches 
\begin{equation} 
\omega_{s,a}({\bf k})=\omega_A\pm J_0-2(J_1\mp J'_1)[cos(k_xa)+cos(k_ya)]. 
\end{equation}  
The antisymmetric branch is almost a dispersion-less, and can be considered as a localized state at each site with energy $\hbar\omega_a\approx\hbar\omega_A+\hbar J_0$. Only symmetric states propagate in the lattice. The symmetric branch band-width is $8\hbar J_1$, between $k=0$ and the boundary of the Brillouin zone at $k=\pi/a$. For small wave vectors, $ka\ll 1$, where $k=|{\bf k}|$, the symmetric branch dispersion reads $\hbar\omega_s(k)=\hbar\omega_A-\hbar J_0-8J_1+\frac{\hbar^2k^2}{2m_{eff}}$, with the effective mass $m_{eff}=\hbar^2/(4J_1a^2)$. At zero wave vector, $k=0$, the symmetric excitation has a shift of $\hbar J_0-8\hbar J_1$ relative to free atoms, which can be easily observed. 
 
As the symmetric coherent exciton states decay approximately with rate $\Gamma_s$, in order to observe exciton effects the excitation line-width needs to be smaller than the exciton band-width, i.e. $\hbar\Gamma_s<8\hbar J_1$. Lets calculate this for Alkali atoms of the transition $^2S_{1/2}- ^2P_{3/2}$, where we have $\hbar\omega_A=1.5-2.5\ eV$ and a line-width of $\hbar\Gamma=1-5\times 10^{-8}\ eV$. The energy transfer $\hbar J(R)$ is calculated from the dipole-dipole interaction between dipoles of $\vec{\mu}^1$ and $\vec{\mu}^2$ separated by a distance $\vec{R}=R\hat{R}$, i.e.
\begin{eqnarray} 
\hbar J(R)&=&\sum_{i,j}\frac{\mu_i^1\mu_j^2}{4\pi\epsilon_0R^3}\left\{-\left(\delta_{ij}-\hat{R}_i\hat{R}_j\right)\ q^2R^2\ \cos qR\right. \nonumber \\ 
&+&\left.\left(\delta_{ij}-3\hat{R}_i\hat{R}_j\right)\left(\cos qR+qR\ \sin qR\right)\right\}, 
\end{eqnarray} 
here $(i,j=x,y,z)$, with $\omega_A=cq$. For linear polarization $\vec{R}=R\hat{x}$ in the $x$ direction $\vec{\mu}^1=\vec{\mu}^2=\mu\hat{x}$, we get $\hbar J_1(R)=\frac{-\mu^2}{2\pi\epsilon_0R^3}\left(\cos qR+qR\ \sin qR\right)$ along $x$. The distance between two atoms at neighbouring sites equals the lattice constant, so that $R=a$. For atoms with $\hbar\omega_A=2\ eV$, a dipole of $\mu=4\ e\AA$ and an optical-lattice of $a=2000\ \AA$, we get $\hbar J_1=10^{-7}\ eV$, which is one order of magnitude larger than the atom line-width so that excitons should be measurable. For the dipole-dipole interaction between two on-site atoms we can neglect the oscillation terms, to get for the previous case $J_0(R)=-\mu^2/(2\pi\epsilon_0R^3)$. For $R=80\ \AA$ we get $\hbar J_0=0.001\ eV$. The next nearest neighbor terms turn out smaller than the excited state line-width and unimportant for exciton formation. 

\ 

Let us now add a planar cavity built of two parallel perfect mirrors to study coherent excitation transfer mediated by the cavity-photons. The optical-lattice is located in the middle between the cavity mirrors as in figure (1). While the lattice lasers are treated as classical fields we consider a quantized cavity field. Opposite to previous work \cite{Ritsch} the cavity-photons energies here are much closer to resonance with the atomic transitions so they will not generate extra periodic forces. 
\begin{figure} 
\centerline{\epsfxsize=5.0cm \epsfbox{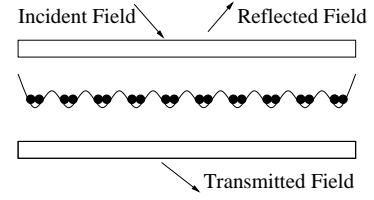}} 
\caption{Optical-lattice with two atoms per site in a cavity.} 
\end{figure} 
 
The electromagnetic field is free in the cavity plane with in-plane wave vector ${\bf k}$ confined only in the perpendicular direction $z$  with wave vector $k_z=m\pi/L$, and $m=0,1,2,\cdots$. Here we consider only the longitudinal mode $m$ closest to resonance with the atom excitation and excited only one polarization. The cavity-photon Hamiltonian thus is given by $H_c=\sum_{\bf k}\hbar\omega_c(k)\ a^{\dagger}_{\bf k}a_{\bf k}$, where $a^{\dagger}_{\bf k},\ a_{\bf k}$ are the creation and annihilation operators of a cavity-photon with in-plane wave vector ${\bf k}$, respectively. The dispersion is $\omega_c(k)=\frac{c}{\sqrt{\epsilon}}\sqrt{k^2+\left(\frac{m\pi}{L}\right)^2}$, with $L$ the distance between the mirrors and $\epsilon=1$. Note that as the optical-lattice is located in the middle between the cavity mirrors $m$ needs to be an odd number to get maximum electric field at the lattice position. 
 
The atomic excitons are coupled to the cavity-photons by the electric dipole interaction $V=-\hat{\mu}\cdot{\bf E}$, where $\hat{\mu}=\vec{\mu}\sum_{i\alpha}(B_i^{\alpha\dagger}+B_i^{\alpha})$ is the material dipole operator, and ${\bf E}$ is the quantized cavity electric field. The transition dipoles $\vec{\mu}$ here are taken to be equal for all the atoms.
The interaction Hamiltonian, in the rotating wave approximation, reads $V=\sum_{{\bf k},i}\sum_{\alpha}\left(f_{{\bf k},i}^{\alpha}\ B_i^{\alpha\dagger}a_{\bf k}+f_{{\bf k},i}^{\alpha\ast}\ a^{\dagger}_{\bf k}B_i^{\alpha}\right)$. The coupling parameter $f_{{\bf k},i}^{\alpha}$ between atom $\alpha$ at site $i$ and a photon with wave vector ${\bf k}$, 
is $f_{{\bf k},i}^{\alpha}=i\sqrt{\frac{\hbar\omega_c(k)}{2LS\epsilon_0}}(\vec{\mu}\cdot\hat{\epsilon}_{\bf k})e^{i{\bf k}\cdot{\bf r}_i^{\alpha}}$, where $S$ is the cavity mirror area, and ${\bf r}_i^{\alpha}$ is the atom position. In terms of symmetric and antisymmetric operators we get $V=\sum_{{\bf k},i,\nu}\left(f_{{\bf k},i}^{\nu}\ B_i^{\nu\dagger}a_{\bf k}+f_{{\bf k},i}^{\nu\ast}\ a^{\dagger}_{\bf k}B_i^{\nu}\right)$, where we defined the symmetric and antisymmetric coupling parameters $f_{{\bf k},i}^{s}=\frac{f_{{\bf k},i}^{1}+f_{{\bf k},i}^{2}}{\sqrt{2}}\ ,\ f_{{\bf k},i}^{a}=\frac{f_{{\bf k},i}^{1}-f_{{\bf k},i}^{2}}{\sqrt{2}}$. As the two atoms are on the same site, we have approximately $f_{{\bf k},i}^{1}\approx f_{{\bf k},i}^{2}$ and the antisymmetric excitations (dark states) almost decouple from the cavity-photons. Only the symmetric excitations can get strongly coupled to the cavity-photons. The position of atom $\alpha$ at site $i$ is ${\bf r}_i^{\alpha}={\bf r}_i+{\bf d}_{\alpha}$. We assumed that each one of the two atoms has the same average on-site position, ${\bf d}_{\alpha}$. For the coupling we can write $f_{{\bf k}}^{\nu}\approx-i\sqrt{\frac{\hbar\omega_c(k)N\mu^2}{4LS\epsilon_0}}\left(1\pm e^{ik|{\bf d}_{2}-{\bf d}_{1}|}\right)$, where $|{\bf d}_{2}-{\bf d}_{1}|$ is the average deviation with a fixed direction. The coupling of the antisymmetric states to the cavity photons, even though weak, can play an important role in coupling these states with the external world. 
 
We get mainly strong coupling cavity photons of transverse wavelength much longer than the distance between the two atoms and excitons of about the same wave vector. Hence we can assume $k|{\bf d}_{2}-{\bf d}_{1}|\ll 1$ and thus we get $f_{{\bf k}}^{s}\equiv f_{k}\approx-i\sqrt{\frac{\hbar\omega_c(k)N\mu^2}{LS\epsilon_0}}\ ,\ f_{{\bf k}}^{a}\approx 0$. In $k$ space we have 
\begin{eqnarray} 
H&=&\sum_{{\bf k}}\left\{\hbar\omega_{s}(k)\ B_{\bf k}^{s\dagger}B_{\bf k}^{s}+\hbar\omega_c(k)\ a^{\dagger}_{\bf k}a_{\bf k}\right. \nonumber \\ 
&+&\left.\hbar f_{{k}}\ B_{\bf k}^{s\dagger}a_{\bf k}+\hbar f_{{k}}^{\ast}\ a^{\dagger}_{\bf k}B_{\bf k}^{s}\right\}+\sum_{i}\hbar\omega_{a}\ B_i^{a\dagger}B_i^{a}. 
\end{eqnarray} 
 
In the strong coupling regime where the exciton and photon line-widths are smaller than the coupling strength, the exciton and the photon are mixed to form the new system excitations, called polaritons \cite{Zoubi,Kavokin} diagonalizing the first part of the above Hamiltonian. As the exciton effective mass is much larger than for photons, the exciton dispersion around the exciton-photon coupling can be neglected and one can simply use $\omega_s\approx\omega_A-J_0-8J_1$.  The diagonalized Hamiltonian is:
\begin{equation} 
H=\sum_{{\bf k}r}\hbar\Omega_r(k)\ A^{\dagger}_{{\bf k}r}A_{{\bf k}r}+\sum_i\hbar\omega_a\ B_i^{a\dagger}B_i^{a}. 
\end{equation} 
We have two polariton branches with dispersions $\Omega_{\pm}(k)=\frac{\omega_{c}(k)+\omega_s}{2}\pm\Delta_k$, where $\Delta_k=\sqrt{\delta_k^2+|f_k|^2}$ and the exciton-photon detuning is $\delta_k=(\omega_{c}(k)-\omega_s)/2$. The splitting at the exciton-photon intersection point, where $\delta_k=0$, is $2|f_k|$, is called the Rabi splitting. The polaritons are coherent superpositions of symmetric excitons and photons, with the diagonal operators $A_{{\bf k}\pm}=X_{k}^{\pm}B_{\bf k}^s+Y_{k}^{\pm}a_{{\bf k}}$ with exciton and photon amplitudes: $X_{k}^{\pm}=\pm\sqrt{\frac{\Delta_k\mp\delta_k}{2\Delta_k}}$, and $Y_{k}^{\pm}=\frac{f_{k}}{\sqrt{2\Delta_k(\Delta_k\mp\delta_k)}}$. 
 
In figure (2.a) we illustrate this plotting the upper and lower polariton energy branches. As example the exciton energy around small in-plane wave vector is taken to be $\hbar\omega_s=1.999\ eV$, and we have $\hbar\omega_a=2.001\ eV$. The distance between the cavity mirrors is $L/m=3102\ \AA$, which is chosen to give zero detuning between the symmetric exciton and the cavity-photon dispersions at zero in-plane wave vector. For $m=3$ we get $L\approx 1\ \mu m$. The transition dipole is $\mu=4\ e\AA$, and the optical-lattice constant is $a=2000\ \AA$. The exciton-photon coupling energy, by using $S=Na^2$, is $|\hbar f|=0.00015\ eV$, where we neglected the $k$ dependence for small in-plane wave vectors. At zero in-plane wave vector the Rabi splitting energy of $0.0003\ eV$ is clear. At large wave vectors the upper polariton branch tends to the cavity-photon dispersion, and the lower branch tends to the symmetric exciton dispersion. In figure (2.b) we plot the exciton and photon weights in the lower and upper polariton branches, $|X^{\pm}|^2$ and $|Y^{\pm}|^2$. At zero in-plane wave vector the polariton is half exciton and half photon. For large in-plane wave vectors the lower polariton becomes much more excitonic than photonic, and vice versa for the upper polariton. 
\begin{figure} 
\centerline{\epsfxsize=4.0cm \epsfbox{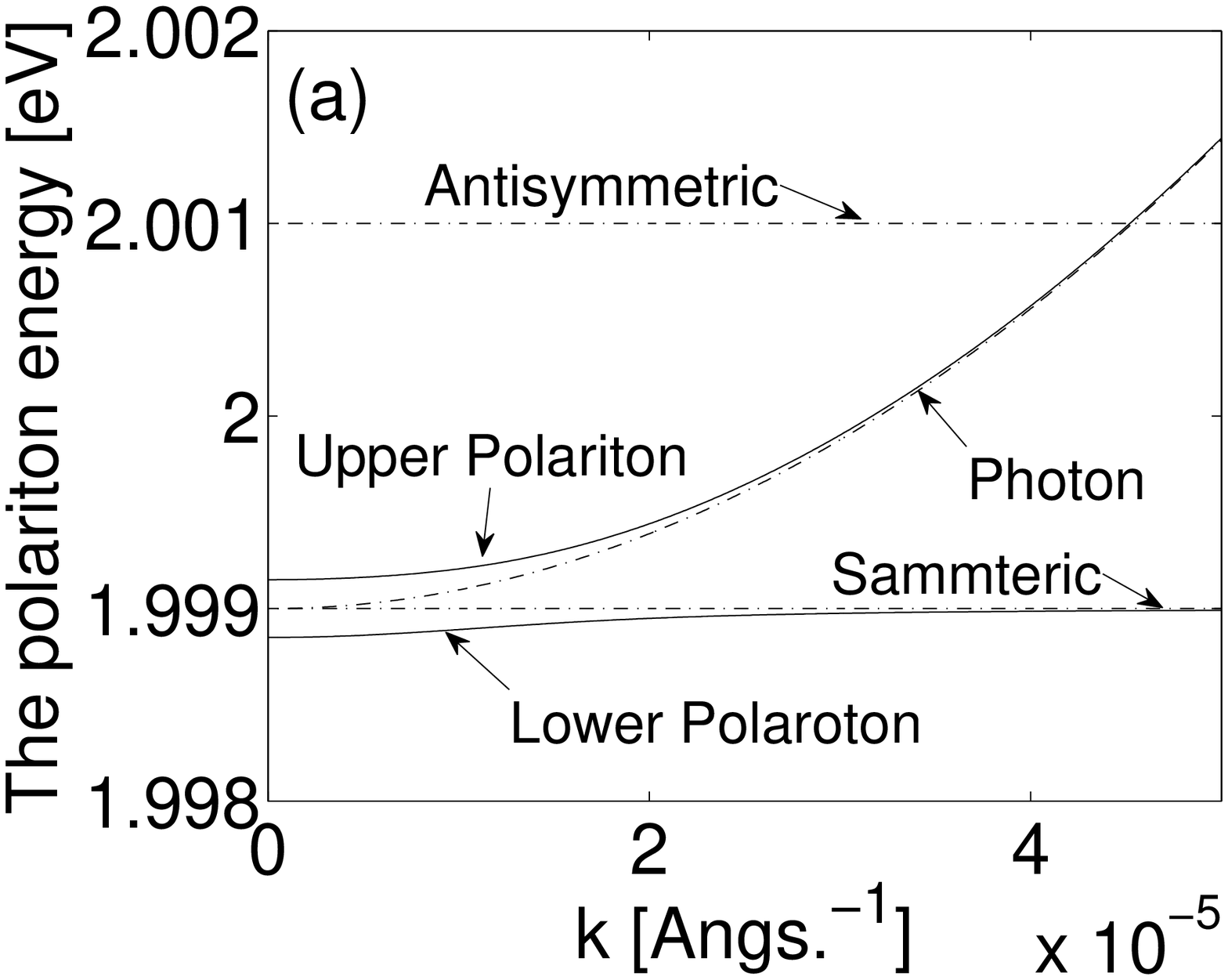}\ \ \ \epsfxsize=4.0cm \epsfbox{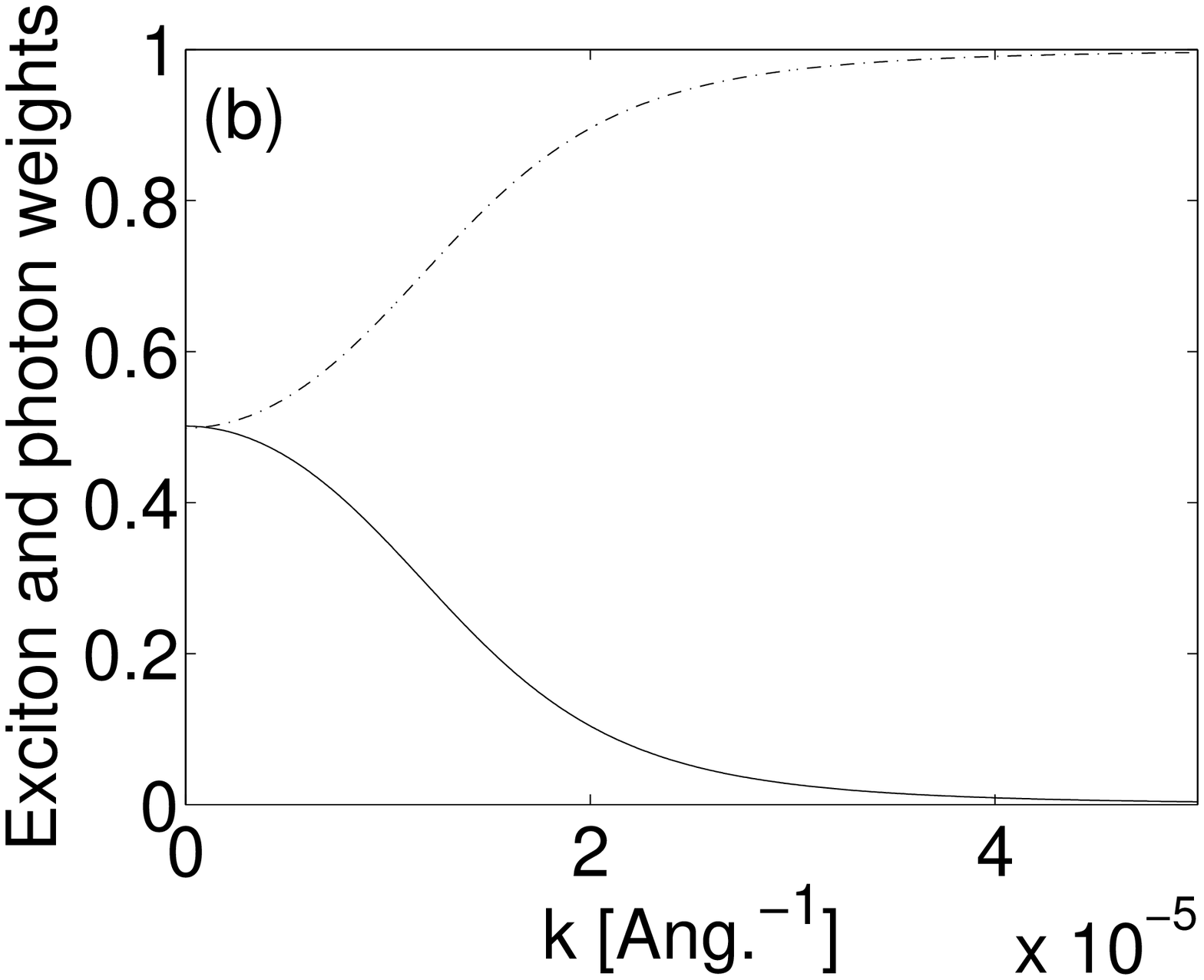}} 
\caption{(a) The upper and the lower polariton branches vs. in-plane wave vector $k$. (b) The symmetric exciton and photon weights vs. in-plane wave vector $k$. In the lower polariton branch the dashed line is for the photon weight, and the full line is for the exciton weight; and vice versa for the upper branch.} 
\end{figure} 

\ 

So far the situation is closely analogous to a lattice with occupancy one with rescaled coupling. However, in order to exhibit and use the extra degrees of freedom here, we consider the case of an optical lattice with asymmetric sites. where one of the orthogonal pairs of counter propagating lasers has a different intensity and thus the on-site potential is elongated in one direction, e.g. $x$. Hence the two on-site atoms will have an average distance $R=|{\bf d}_{2}-{\bf d}_{1}|$ at the $x$ direction, even after considering the local atom wave functions. Thus we have $R\ll a$  and then $J_0\gg J_1$. The atomic transition dipole is induce by the cavity photon, and is oriented along the electric field of the cavity photon. We treat the case of in-plane transition dipoles, namely the general dipole is $\vec{\mu}=\mu(\cos\theta,\sin\theta)$, which is depend on $\theta$, the angle between the dipole $\vec{\mu}$ and the $x$ axis. The resonance dipole-dipole interaction between the two on-site atoms is $\hbar J_0(\theta)=\frac{\mu^2}{4\pi\epsilon_0R^3}\left(1-3\cos^2\theta\right)$. We have $\hbar J_0(\theta=0)=-2\mu^2/(4\pi\epsilon_0R^3)$, and $\hbar J_0(\theta=90)=\mu^2/(4\pi\epsilon_0R^3)$, with $\hbar J_0(\theta\approx 54.74)=0$. The detuning energy between symmetric and antisymmetric states can change sign and also for a fixed polarization can be vanished. 

Using the above results, in figure (3.a) we plot the symmetric and antisymmetric states, and also the two polariton branches, as a function of $\theta$, for the case of zero in-plane wave vector, that is $k=0$. It is clear that the symmetric-antisymmetric states change sign at $\theta\approx 54.74$. At this angle the two states are degenerate. We chose zero detuning between the cavity photon and the antisymmetric state at $\theta=0$. The maximum detuning between the symmetric state and the cavity photon appears at $\theta=90$. In figures (3.b) we plot the excitonic and photonic weights in the lower and upper polariton branches as a function of the angle $\theta$. The symmetric-antisymmetric splitting and the Rabi splitting are easily controlled by changing the field polarization direction. 
\begin{figure} 
\centerline{\epsfxsize=4.0cm \epsfbox{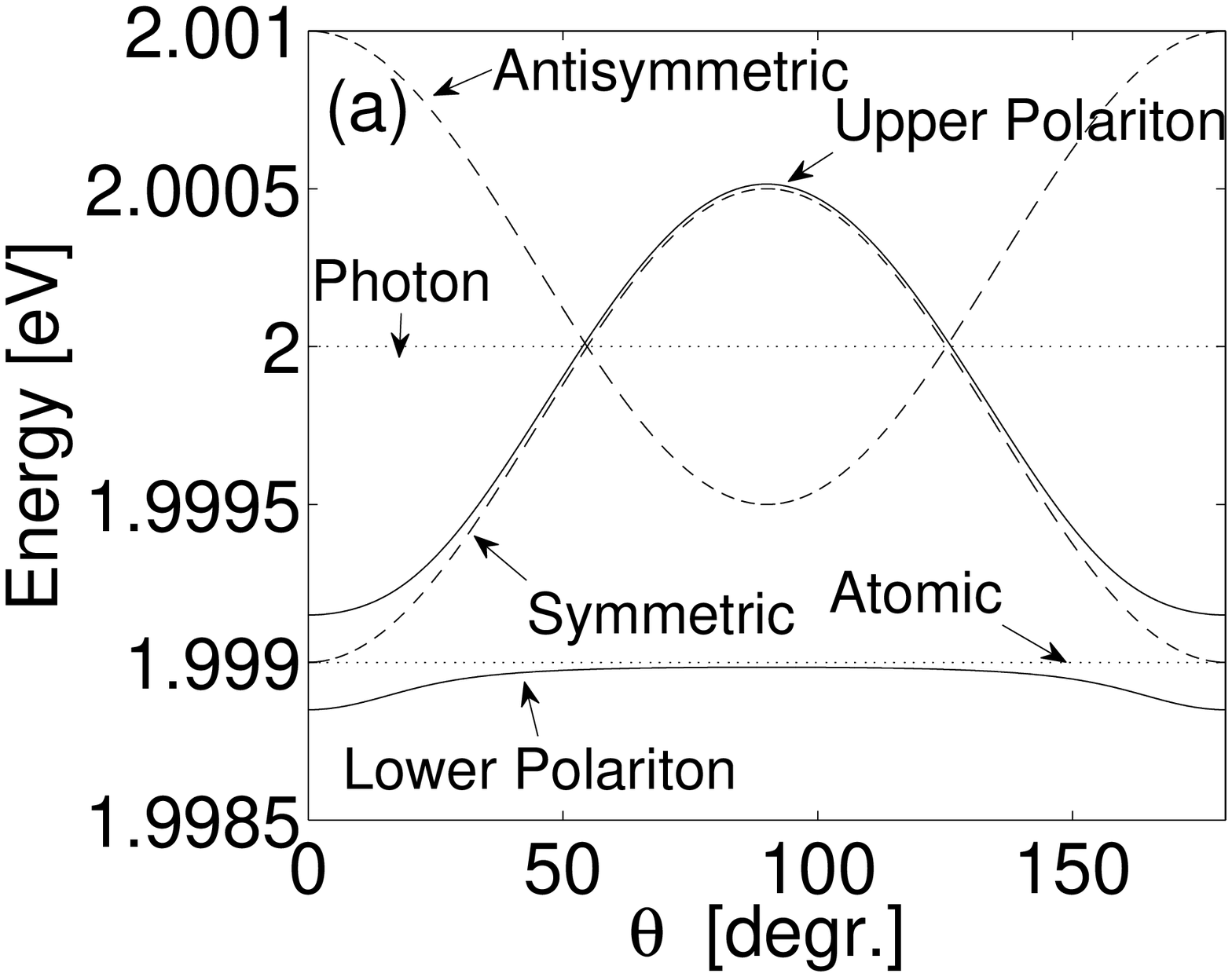}\ \ \ \epsfxsize=4.0cm \epsfbox{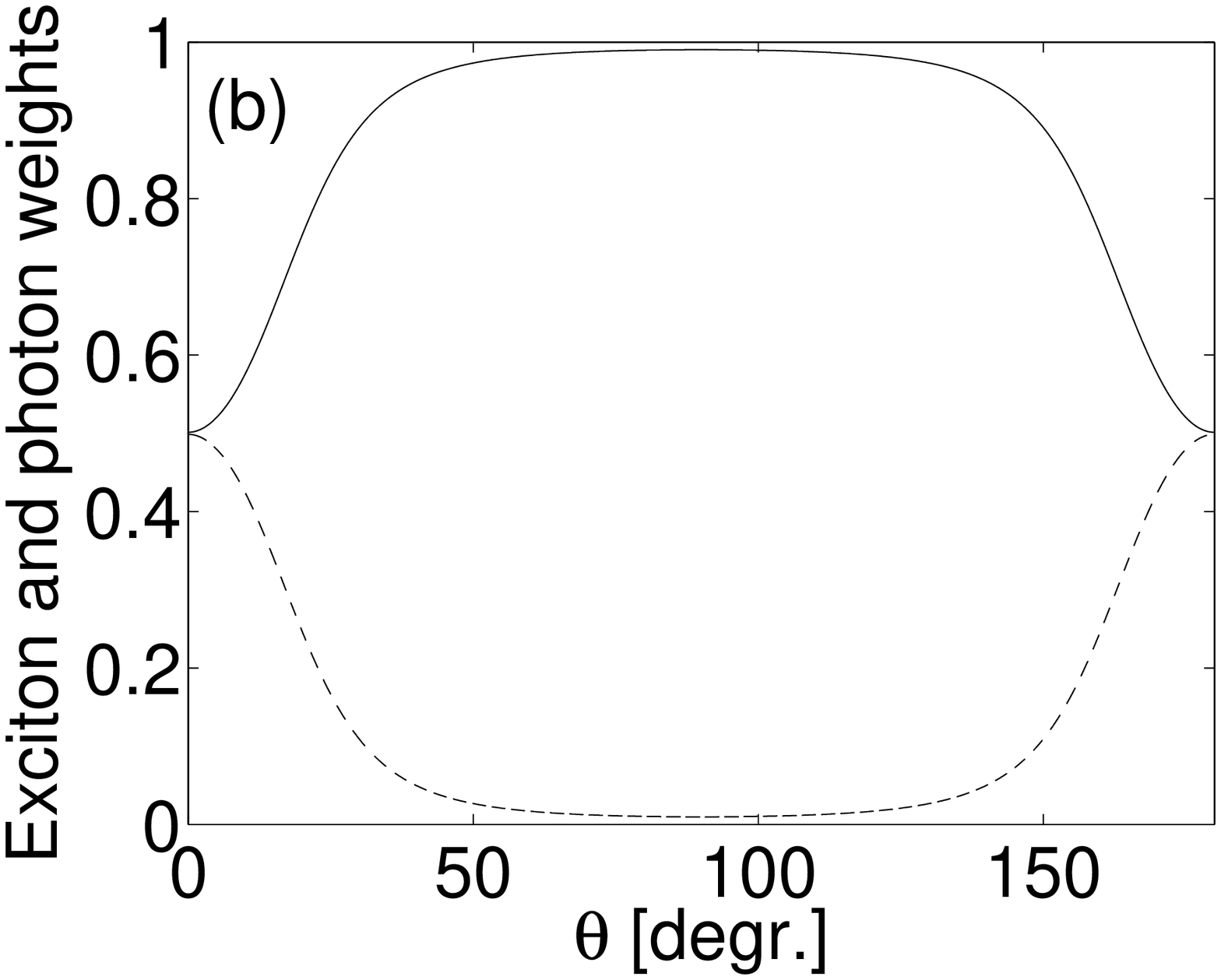}} 
\caption{(a) The symmetric and antisymmetric states, and the upper and the lower polariton branches vs. the angle $\theta$, for the $k=0$ case. (b) The symmetric exciton and photon weights vs. the angle $\theta$. In the lower polariton branch the dashed line is for the photon weight, and the full line is for the exciton weight; and vice versa for the upper branch.} 
\end{figure} 

In order to observe these system eigenmodes we couple the internal cavity modes to the external world, i.e. the external radiation field in and out coupled through the cavity mirrors, and calculate the cavity input and output fields in a standard quantum optical approach. Similarly we will include atomic spontaneous emission via an effective exciton damping. In figure (4) we plot the corresponding transmission and reflection spectra \cite{Hashem} for an incident field with zero in-plane wave vector, $k=0$, where the electric field is parallel to the mirrors. We choose the following numbers for the line widths, the symmetric exciton line width is $\hbar\Gamma_s=10^{-7}\ eV$, the mirror line widths is $\hbar\gamma=10^{-5}\ eV$. In figure (4.a) we plot the transmission, and in figure (4.b) the reflection, for different polarization angles. The peaks of the transmission, and the dips of the reflection correspond to the two polariton branches, which are the real eigenmodes of the system. Large transmission and reflection in regions where the polariton is much more photonic than excitonic is obtained.
\begin{figure}
\centerline{\epsfxsize=4.0cm \epsfbox{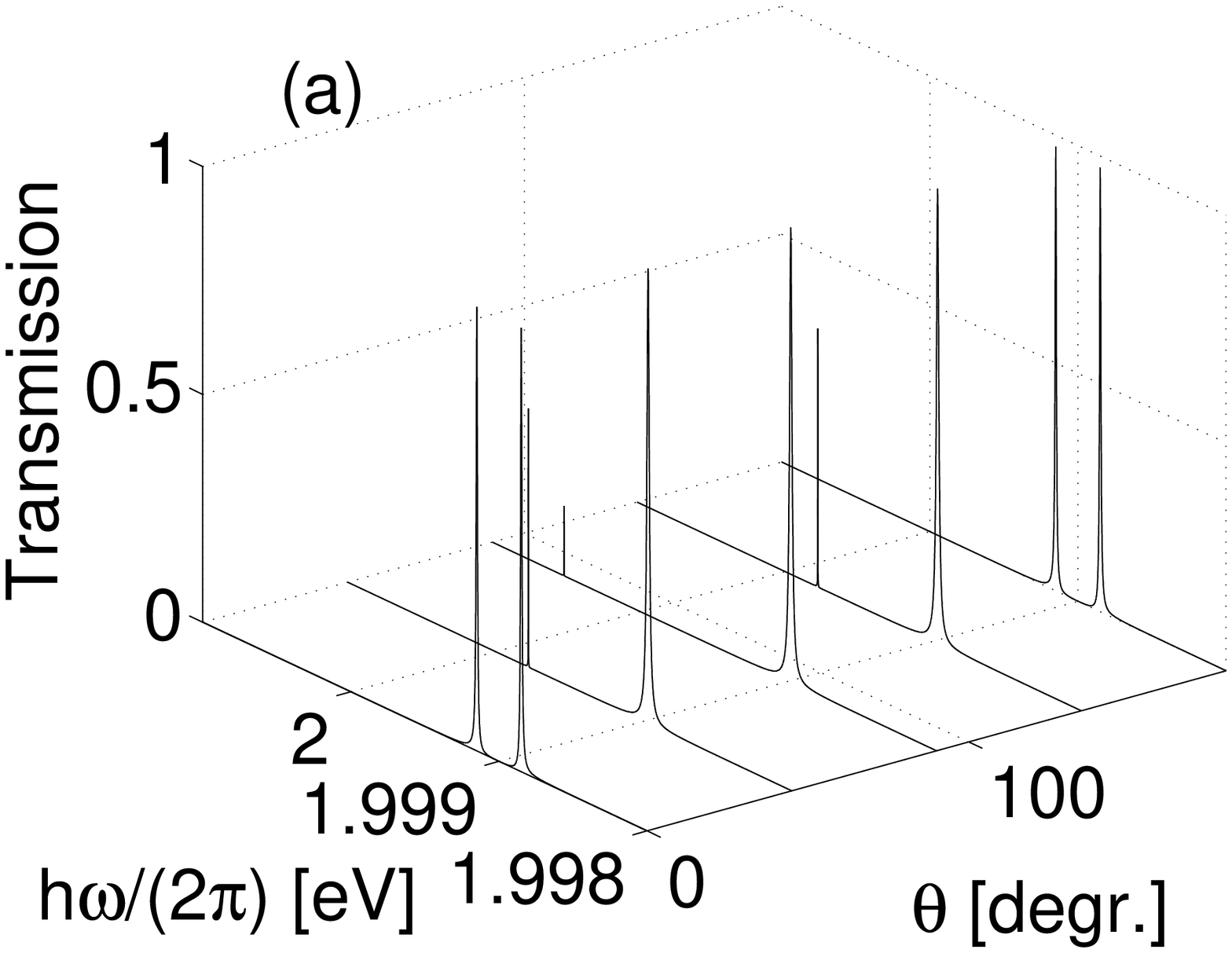}\ \epsfxsize=4.0cm \epsfbox{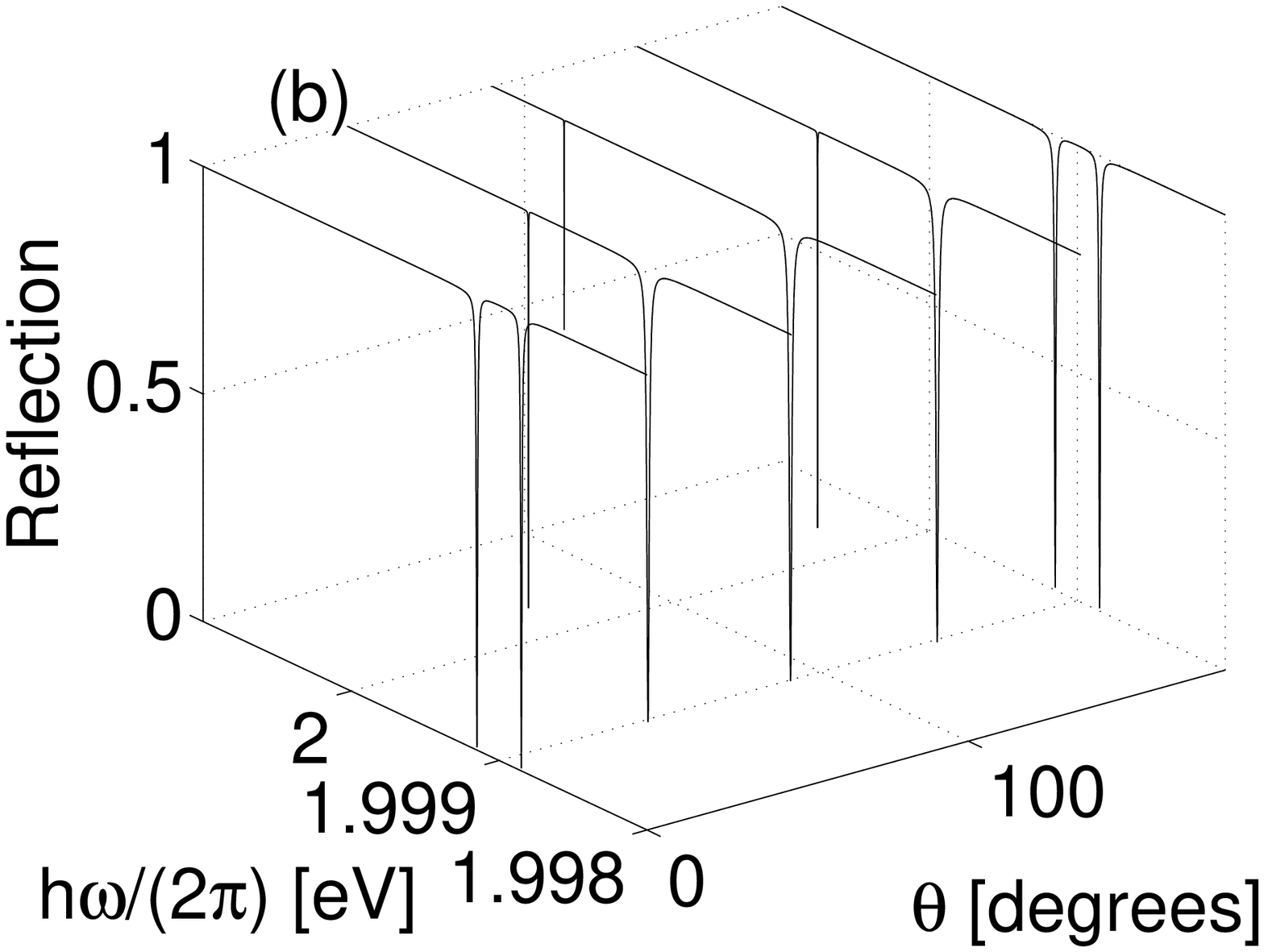}}
\caption{(a) The transmission spectrum at zero in-plane wave vector, for different angles $\theta$. (b) The reflection spectrum at zero in-plane wave vector, for different angles $\theta$.}
\end{figure}

\ 

In a summary excitons and cavity polaritons involving resonant excitations of ultracold-atoms in an optical-lattice acquire intriguing new properties for the case of two atoms trapped at each site. Antisymmetric local excitations with a long lifetime and weak neighbor coupling point to applications for long time memory in optoelectronics and quantum information. Symmetric excitations strongly couple to neighboring atoms and cavity photons and those form polaritons mediating controlled long range interactions. Controlled excitations can be facilitated via decay of higher energy states by nonlinear optical processes. Similarly one could exploit the weak coupling of the antisymmetric states to the cavity photons in order to write and read information of these states, while the symmetric state polaritons can be used as diagnosis and readout tool.

The work was supported by the Austrian Science Fund (FWF), through the Lise-Meitner Program (M977).

\end{document}